\documentclass[%
 reprint,
superscriptaddress,
showpacs,
amsmath,amssymb,
 aps
]{revtex4-1}

\usepackage{graphicx}
\usepackage{dcolumn}
\usepackage{bm}
\usepackage{color}
\usepackage{here}

\usepackage{url,listings}

\begin{document}
\title{Boosting Materials Modeling by using Game Tree Search}
\author {Ryohto Sawada}
\affiliation{IoT Devices Research Laboratories, NEC Corporation, Tsukuba 305-8501, Japan}
\author{Yuma Iwasaki}
\affiliation{IoT Devices Research Laboratories, NEC Corporation, Tsukuba 305-8501, Japan}
\affiliation{JST, PRESTO, Saitama, 332-0012, Japan}
\author{Masahiko Ishida}
\affiliation{IoT Devices Research Laboratories, NEC Corporation, Tsukuba 305-8501, Japan}

\begin{abstract}
We demonstrate a heuristic optimization algorithm based on the game tree search for multi-component materials design. The algorithm searches for the largest spin polarization of seven-component Heusler alloys. The algorithm can find the peaks quickly and is more robust against local optima than Bayesian optimization approaches using the expected improvement or upper confidence bound approaches. We also investigate Heusler alloys including anti-site disorder and \textcolor{black}{show that [Fe$_{0.9}$Co$_{0.1}$]$_{2}$Cr$_{0.95}$Mn$_{0.05}$Si$_{0.3}$Ge$_{0.7}$ has the potential to be a high spin polarized material with robustness against anti-site disorder.}
\end{abstract}

\maketitle

The complexity of industrial materials is increasing as a result of technological progress in materials processing. However, optimization of materials is affected by the curse of dimensionality; the difficulty increases exponentially with the number of parameters (e.g., number of components and heat treatment conditions) \cite{bellman2003}. For this reason, efficient search algorithms that find optimum parameters by operating on only a few sampling points are in great demand to decrease costs.

A well-known search strategy is to determine the next sampling point according to the previous results. One popular algorithm adopting this strategy is the genetic algorithm (GA) \cite{goldberg1989, poli2008, hart2005}. Previous studies have shown that the GA can optimize castings \cite{darby2002, santos2003, castro2004, anjidan2006} and magnetic alloys \cite{blum2005, hart2005}. However, it is also reported that controlling the genes' diversity is so difficult that the algorithm usually converges prematurely and induces wasteful duplication of sampling points \cite{gen1999, yang2016}. To decrease the number of redundant sampling points, not only the expected result but also the expected uncertainty should be considered prior to selecting the sampling points.

\begin{figure}[tb]
\includegraphics[width=8cm, bb = 0 0 550 780]{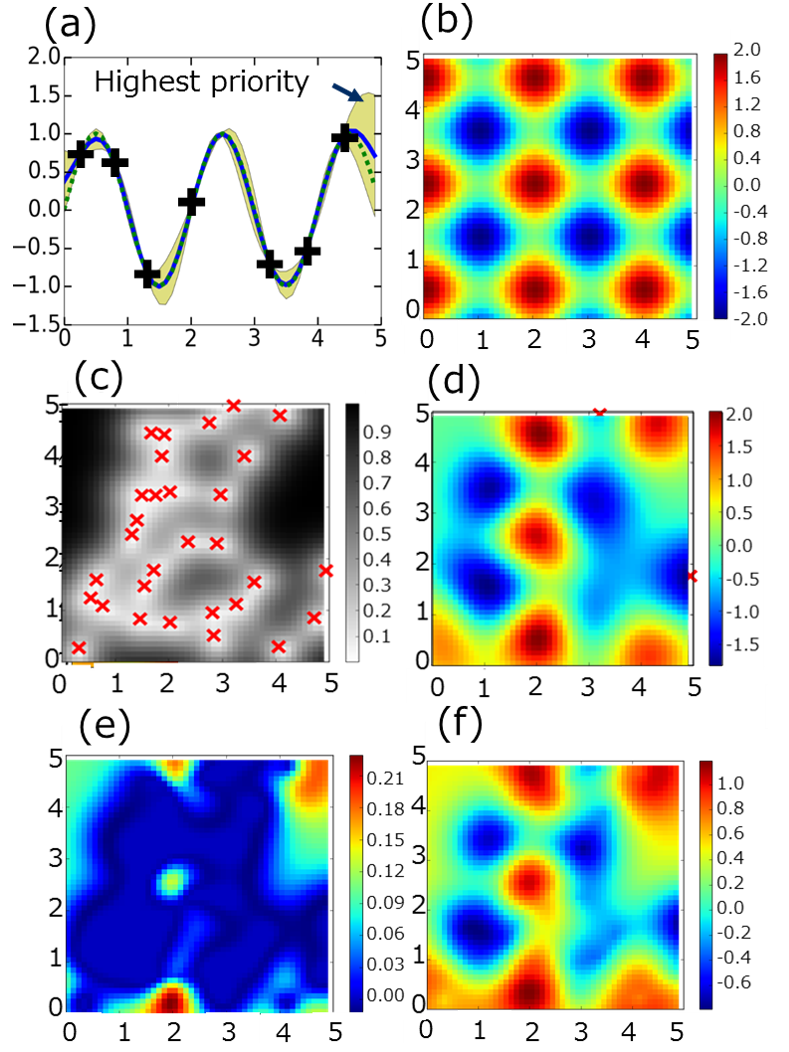}
\caption{Example of Gaussian process regression in one-dimensional (a) and two-dimensional (b-f) cases. In (a), the green dashed line is the exact value (sine function), the black crosses are sampling points, the blue solid line is the expected result, and the yellow area is the expected error. The next sampling point is indicated by the black arrow. (b) is the exact value, (c) is the expected error, where the crosses are sampling points and the number of the data is 30, (d) is the expected result, and (e) and (f) are the priorities $P(x)$ obtained by Eq. \ref{eq:ei} and Eq. \ref{eq:ucb}, respectively, at $C$ = 1.0.}
\label{fig:gp}
\end{figure}

Fig. \ref{fig:gp}(a) shows an example of Gaussian process regression, which is a useful way to take into account both the expected result and uncertainty. In this plot, the black crosses are sampling points and the green dashed line shows the exact value. The blue solid line and yellow-shaded area are, respectively, the expected result and range of uncertainty estimated from the previous results. The next sampling point (indicated by the black arrow) is determined in accordance with the priority $P(x)$, e.g.,
\begin{equation}
P(x) = \int_{f_{max}}^{\infty} dy (y-f_{max}) \exp( -(f(x) - y)^2 /\sigma(x)) 
\label{eq:ei}
\end{equation}
or
\begin{equation}
P(x) = f(x) + C\sigma(x) ,
\label{eq:ucb}
\end{equation}
where $f(x)$ is the expected value, $\sigma$ is the expected error, $f_{max}$ is the best (maximum) result obtained by the previous results, and $C$ is a hyperparameter indicating the weight of ambiguity. Equation \ref{eq:ei} is referred to as the expected improvement (EI) algorithm \cite{jones1998}, and Eq. \ref{eq:ucb} is called the upper confidence bound (UCB) strategy \cite{auer2003}. The EI algorithm and UCB strategy are simple and have been used in materials modeling of low-degree-of-freedom systems \cite{okamoto2017, srinivas2010}.

However, the EI algorithm and UCB strategy are hardly applicable to multi-dimensional optimization for two reasons. The first is that the cost of calculating the expected values and errors in the entire search space exponentially increase with the number of parameters and resolution. The second is that these approaches are vulnerable to incorrect predictions. This vulnerability can be seen in the example of Gaussian process regression for the two-dimensional function shown in Figs. \ref{fig:gp} (b-d). Figure \ref{fig:gp} (b) is the exact value, (c) is the expected error (the red crosses are sampling points), and (d) is the expected value. One can see that Gaussian process regression makes incorrect predictions around (0, 4.5), (0, 2.5), and (4, 2.5) [Fig. \ref{fig:gp} (d)] and most of the search space has a large error, unlike the one-dimensional case [Fig. \ref{fig:gp} (c)]. Figures \ref{fig:gp} (e) and (f) show the priorities obtained by Eq. \ref{eq:ei} and Eq. \ref{eq:ucb}, respectively. One can see that $P(x)$ around the overlooked peaks is too low for any of them to be selected as the next sampling point. In this case, $P(x)$ around the overlooked peaks can be raised by tuning $C$; however, the appropriate value of $C$ depends strongly on the target function and the previous results, and it varies during the optimization. Therefore, it is difficult for the EI algorithm and UCB strategy to avoid local optima.

We addressed these problems by using a game tree search. The game tree search can manage the spatial resolution of the expectation by varying the depth of the tree. It maintains a balance between a dense search (optimization around a peak) and a sparse search (exploration for unknown peaks), and we found that it is about nine times faster than previous methods at optimizing the spin polarization of multi-component Heusler alloys.

\begin{figure}[tb]
\includegraphics[width=8cm, bb = 0 0 950 760]{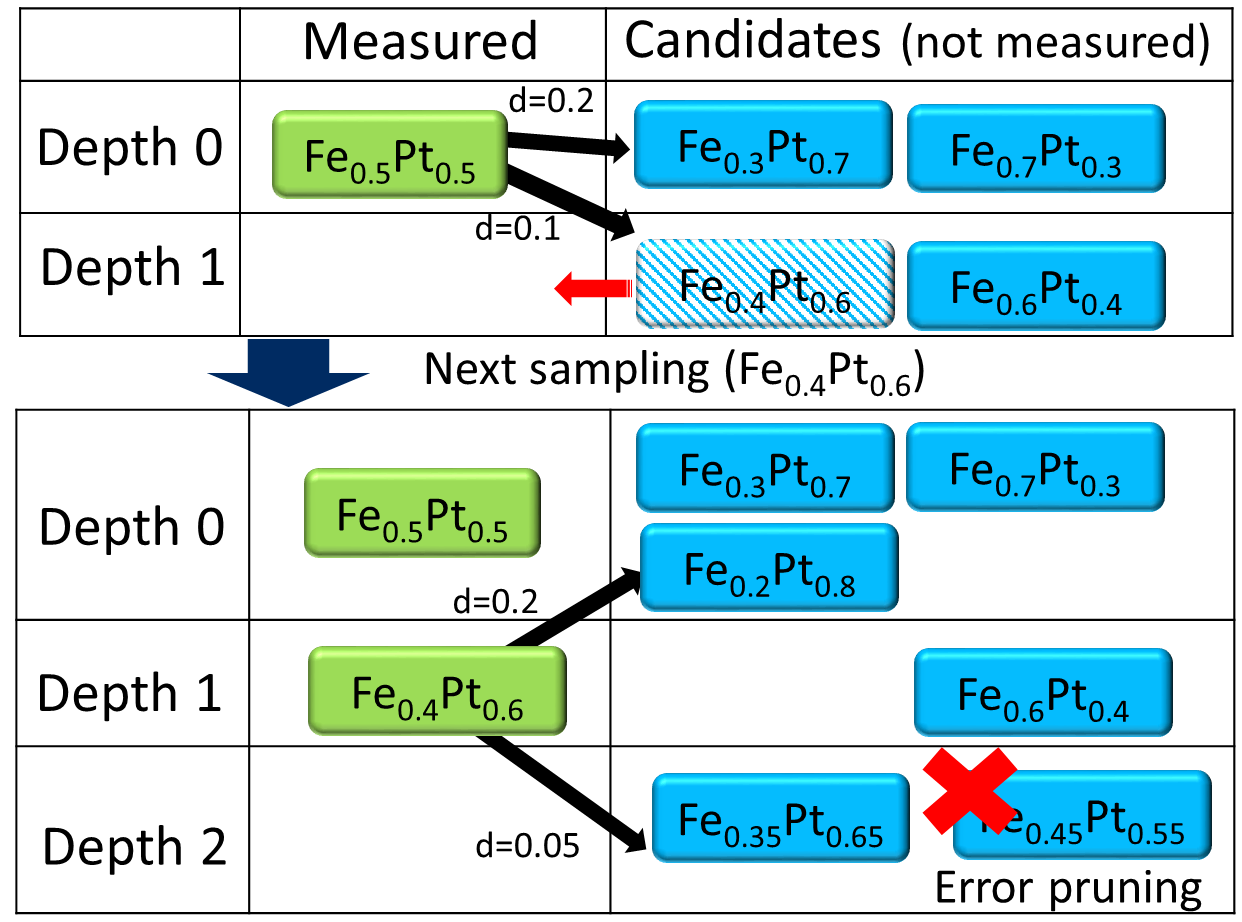}
\caption{Schematic image of game tree search.}
\label{fig:tree}
\end{figure}

Figure \ref{fig:tree} shows a virtual sampling in which a game tree search was used to optimize the composition of Fe$_{x}$Pt$_{1-x }$ alloy with regard to a certain physical value, i.e., the magnetic moment. The game tree search limits the candidates for the next sampling points such that they are only in the vicinity of the current sampling point and sets the spatial resolution in accordance with the depth of the tree. The distance between the candidates and the sampling point takes two kinds of value; $d_{0}$ and $2^{-D}d_{0}$, where $d_{0}$ is the initial spatial resolution and $D$ is the depth of the point. In the upper panel of Fig. \ref{fig:tree}, Fe$_{0.3}$Pt$_{0.7}$, Fe$_{0.7}$Pt$_{0.3}$, Fe$_{0.6}$Pt$_{0.4}$, and Fe$_{0.4}$Pt$_{0.6}$ are generated from Fe$_{0.5}$Pt$_{0.5}$, where $d_{0}$ is set to be 0.2. The next sampling point is determined by comparing the priorities of the candidates by using Eq. \ref{eq:ei} or \ref{eq:ucb} (Fe$_{0.4}$Pt$_{0.6}$ is selected in the example). After the measurement, the game tree generates the candidates from the current set of measurement points (lower panel of Fig. \ref{fig:tree}). If the estimated uncertainty is lower than $e_{min}$ or the estimated result is lower than $r_{min} v_{best}$, we can exclude this point from the set of candidates (error pruning); $e_{min}$ and $r_{min}$ are parameters set by the user, while $v_{best}$ is the best value among the previous measurements. Error pruning helps to avoid redundant measurements and accelerates convergence. The pseudo code of the game tree search is shown in Listing \ref{pcode}. 

\begin{lstlisting}[tabsize=3, caption=Pseudo code of the game tree search.,label=pcode]
class Leaf :
	depth  /* depth of leaf */
	comp  /* composition */

Function Mutation(parent, add_depth):
	if add_depth == true:
		leaf.depth = parent.depth + 1
	else:
  		leaf.depth = 0
 	/* norm of random_vector is set to be 1 */
	leaf.comp = parent.comp + d0 * (0.5^leaf.depth) * random_vector
	return leaf

Function GameTree :
	Table <Leaf> gametree 
	ucbtree.add(startpoint)
	best = -INF
	do loop:
		maxp = -INF
		for each leaf in leafs:
			v = Priority(leaf) /* UCB or EI */
			if maxp < v:
				maxleaf = leaf 
				maxp = v
		result = Measurement(maxleaf)
		gametree.delete(maxleaf)
		if result > best :
			best = result
		rb = result / best
		leaf1 = Mutation(maxleaf, false)
		leaf2 = Mutation(maxleaf, true)
		/* error pruning */
		if result/best > rmin and Error(leaf1) > emin: 
			gametree.add(leaf1)
		if result/best > rmin and Error(leaf2) > emin:
			gametree.add(leaf2)
\end{lstlisting}

\begin{figure}[tbp]
\includegraphics[width=8cm, bb = 0 0 580 830]{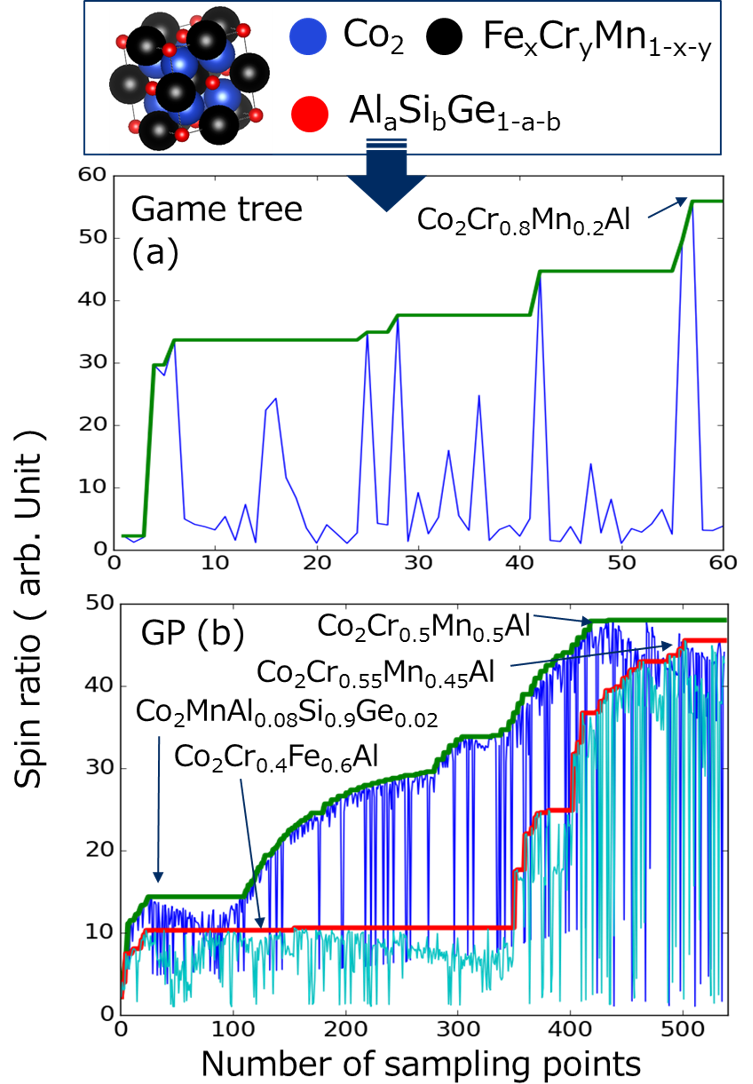}
\caption{Results of optimization of spin polarization of Co$_{2}$Cr$_{x}$Mn$_{y}$Fe$_{1-x-y}$Al$_{a}$Si$_{b}$Ge$_{1-a-b}$. The x-axis shows the number of sampling points, thin lines show the spin ratio of the x-th sampling point, and the bold lines show the maximum spin ratio of the first to x-th sampling points. (a) Results of game tree search. (b) Results of EI (green solid line) algorithm and UCB strategy (red solid line).}
\label{fig:comparison}
\end{figure}

Now we demonstrate a four-dimensional case of composition optimization of Heulser alloys, materials that are potentially useful in random access memories and spin transfer devices \cite{webster1969, galanakis2005}. To find promising compositions, first-principles simulations have been used because of their low cost \cite{groot1983, fujii1990, tobola2001, galanakis2003, block2004, stopa2006}. However, Heusler alloys have too many combinations to examine them all. Here, we optimized the spin polarization $p$ of Co$_{2}$Cr$_{x}$Mn$_{y}$Fe$_{1-x-y}$Al$_{a}$Si$_{b}$Ge$_{1-a-b}$ by using a game tree search, the EI algorithm, and the UCB strategy. The spin polarization was defined as (n$\uparrow$-n$\downarrow$)/(n$\uparrow$+n$\downarrow$), where n$\uparrow$ and n$\downarrow$ are the respective density-of-state values of up and down spin electrons at the Fermi energy. The density of states was calculated using the Korringa Kohn Rostoker (KKR) band structure and coherent potential approximation (KKR-CPA method) \cite{akai1982, ogura2007} with the AKAI-KKR package \cite{akai1982}. The crystal structure was assumed to be full-Heusler [inset of Fig. \ref{fig:comparison}], and the lattice constant was made to minimize the total energy \textcolor{black}{in every iteration}. The priority of the candidates in the game tree search was evaluated using Eq. \ref{eq:ucb}. The importance of the ambiguity $C$ was set to be the same as in \cite{auer2003} for the UCB strategy and 1.0 for the game tree search. The parameters of the game tree were d$_{0}$ = 0.8, $r_{min}$ = 0.1, and $e_{min}$ = 0.1. The first sampling composition was $x = y = a = b = 0.33$. We regarded the distance between components $D(x,y,a,b; x',y',a',b')$ as the Euclidean distance of the normalized components,
\begin{eqnarray}
D(x,y,&a&,b; x',y',a',b')^{2} = \nonumber \\
&\ & (x-x')^{2} + (y-y')^{2} + (x+y-x'-y')^{2} \nonumber \\
&+& (a-a')^{2} + (b-b')^{2} + (a+b-a'-b')^{2} .
\label{eq:eu}
\end{eqnarray}
We used the spin ratio n$\uparrow$/n$\downarrow$ as the result of each measurement instead of the spin polarization. The spin ratio monotonically increases with respect to the spin polarization, and it is useful for accelerating convergence around $p \approx 1$. Figure \ref{fig:comparison}(a,b) shows the calculated spin ratio (y-axis) and the measurement number (x-axis) when using the game tree search (a) and EI algorithm and UCB strategy (b). In this case, the game tree search reached the expectation that Co$_{2}$Cr$_{0.8}$Mn$_{0.2}$Al has the largest spin ratio. \textcolor{black}{This expectation does not contradict previous theoretical studies}\cite{galanakis2002, galanakis2004}. On the other hand, the EI algorithm and UCB strategy both get trapped in local optima around Co$_{2}$Cr$_{0.5}$Mn$_{0.5}$Al, despite requiring nine times more sampling points than the game tree search needed. In particular, the EI algorithm and UCB strategy spent a lot of time escaping from local optima, e.g., Co$_{2}$MnAl$_{0.08}$Si$_{0.9}$Ge$_{0.02}$ and Co$_{2}$Cr$_{0.4}$Fe$_{0.6}$Al. This problem stems from that Gaussian process regression made incorrect predictions during the first several steps because of the few sampling points that were available to it. On the other hand, the game tree search escaped from local optima quickly. It limited the resolution of the sampling by using the tree depth. This limitation forced it to measure compositions outside the local maximums. Once a higher peak was found, the candidates around the local optima were pruned. 

We also examined a more practical case. Anti-site disorder is inevitable in actual Heusler alloys. Therefore, the effect of anti-site disorder should be considered in order to predict actual materials. It can be estimated by calculating the band gap around the Fermi energy \cite{hulsen2009, choudhary2016} and by calculating the change in spin polarization as a result of swapping atoms \cite{miura2004, picozzi2004, galanakis2007, ouardi2010, kudrnovsky2013}. 
\begin{figure}[H]
\includegraphics[width=8cm, bb = 0 0 550 420]{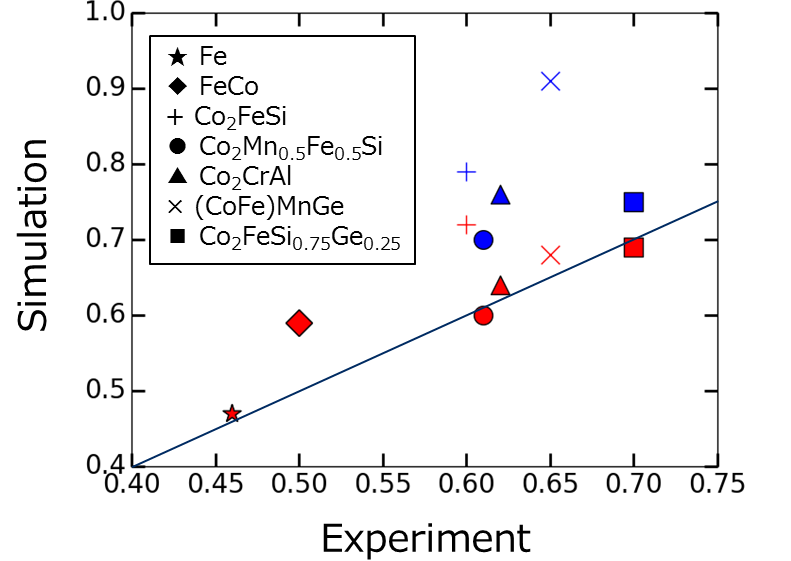}
\caption{Comparison of spin polarizations determined in \cite{sato2015} (x-axis) and from a KKR-CPA calculation (y-axis), where 0(blue)/10(red) percent of Y atoms are swapped with X atoms for X$_{2}$YZ Heusler alloys.}
\label{fig:table}
\end{figure}
Figure \ref{fig:table} compares the spin polarizations of X$_{2}$YZ Heusler alloys determined by the method of \cite{sato2015} and a KKR-CPA calculation. One can see that swapping atoms can eliminate the effect of anti-site disorder.

\begin{figure}[H]
\includegraphics[width=8cm, bb = 0 0 600 630]{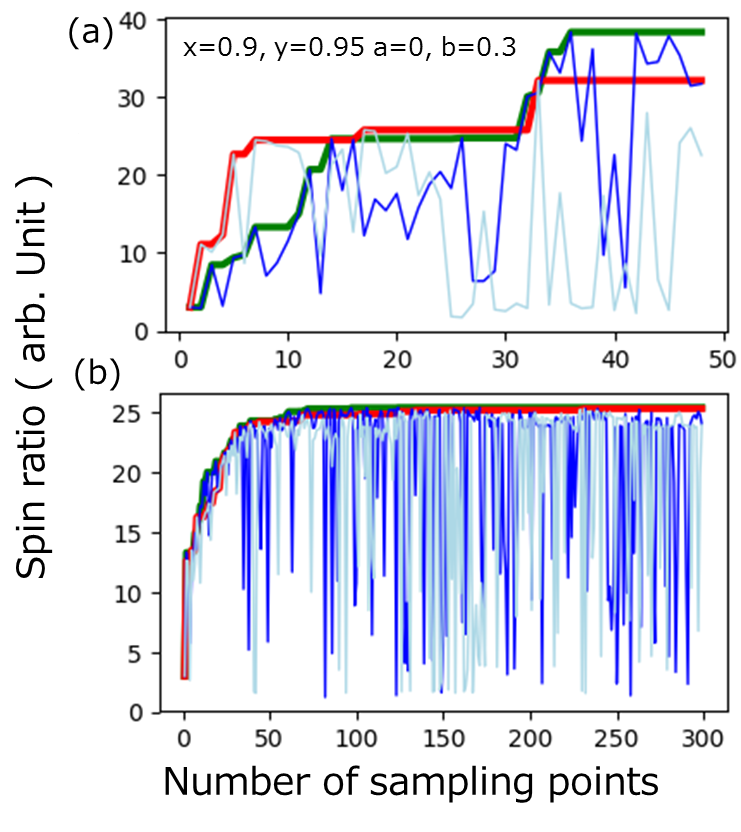}
\caption{Results of optimizing the spin polarization of  [[Fe$_{x}$Co$_{1-x}$]$_{0.975}$ [Cr$_{y}$Mn$_{1-y}$]$_{0.025}$]$_{2}$ [[Cr$_{y}$Mn$_{1-y}$]$_{0.95}$ [Fe$_{x}$Co$_{1-x}$]$_{0.05}$] [Al$_{a}$Si$_{b}$Ge$_{1-a-b}$] by using (a) game tree search where the priorities of the candidates is defined by Eq. \ref{eq:ucb} (red solid line) and Eq. \ref{eq:ei} (green solid line), (b) EI (green solid line) algorithm and UCB strategy (red solid line). The x-axis shows the number of sampling points, the thin line shows the spin ratio of the x-th sampling point, and the bold line shows the maximum spin ratio of the first to x-th sampling points.}
\label{fig:disorder}
\end{figure}

We optimized the spin polarization of [[Fe$_{x}$Co$_{1-x}$]$_{0.975}$ [Cr$_{y}$Mn$_{1-y}$]$_{0.025}$]$_{2}$ [[Cr$_{y}$Mn$_{1-y}$]$_{1-\alpha/2}$ [Fe$_{x}$Co$_{1-x}$]$_{\alpha}$] [Al$_{a}$Si$_{b}$Ge$_{1-a-b}$], where $\alpha$ percent of the [Cr$_{y}$Mn$_{1-y}$] was swapped with [Fe$_{x}$Co$_{1-x}$]. We fixed the percentage of anti-site disorder ($\alpha = 0.05$) and \textcolor{black}{allowed the dopant to fill both the (0,0,0) and (1/2, 1/2, 1/2) positions equally}, because optimizing the disorder conditions would have required a huge amount of computational resources. The game tree search can also be used to optimize the disorder; this issue will be addressed in the future. Figure \ref{fig:disorder} shows the results. We found that $x = 0.9$, $y = 0.95$, $a = 0$, $b = 0.3$ had the largest spin ratio.

\textcolor{black}{We examined the robustness of the spin polarization of this composition by changing $\alpha$ to 0.4. The spin ratio was 6.6 (spin polarization of 0.74), which is higher than that of Co$_{2}$MnSi\cite{picozzi2004}. The origin of the reduction in spin polarization is thought to be the minority energy gap arising from the anti-site disorder. \cite{picozzi2004, galanakis2006}. Modulation of the energy gap by doping is theoretically possible, but practically difficult, because how doping affects the energy gap is difficult to predict. Our implementation will open the way to boosting practical optimizations like this.}

In conclusion, we developed a game tree search algorithm for multi-dimensional optimization. Unlike previous methods, the game tree search is robust against local optima because the resolution of the search can be controlled in accordance with the depth of the tree and local optima can be pruned. We demonstrated that it is about nine times faster at optimizing the spin polarization of multi-component Heusler alloys than the EI algorithm or the UCB strategy. \textcolor{black}{We also found that [Fe$_{0.9}$Co$_{0.1}$]$_{2}$Cr$_{0.95}$Mn$_{0.05}$Si$_{0.3}$Ge$_{0.7}$ has the potential to be a high spin polarized material with robustness against anti-site disorder}. The algorithm is applicable not only to composition optimization, but also to a wide range of topics where regression usually fails due to unexpected characteristics inside real materials. The present implementation will open the way to boosting materials development with AI algorithms.

This work was financially supported by a JST-ERATO Grant (Number JPMJER1402) and JST-PRESTO Grant ( JPMJPR17N4).

\section{Appendix}
\begin{figure}[h]
\includegraphics[width=7cm, bb = 0 0 420 425]{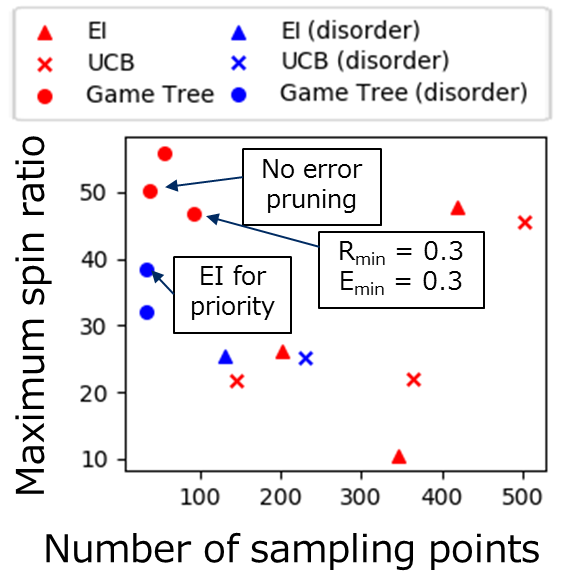}
\caption{Summary of statistics of the optimization of Co$_{2}$Cr$_{x}$Mn$_{y}$Fe$_{1-x-y}$Al$_{a}$Si$_{b}$Ge$_{1-a-b}$ and [[Fe$_{x}$Co$_{1-x}$]$_{0.975}$ [Cr$_{y}$Mn$_{1-y}$]$_{0.025}$]$_{2}$ [[Cr$_{y}$Mn$_{1-y}$]$_{0.95}$ [Fe$_{x}$Co$_{1-x}$]$_{0.05}$] [Al$_{a}$Si$_{b}$Ge$_{1-a-b}$] (labeled "disorder"). The x-axis shows the number of samplings needed to reach the maximum spin ratio, and the y-axis shows the maximum spin ratio. The simulation conditions are the same as those used to make the plots in Fig. \ref{fig:comparison} and Fig. \ref{fig:disorder}.}
\label{fig:sum}
\end{figure}

\textcolor{black}{We repeated the simulations and summarized the results in Fig. \ref{fig:sum}. The efficiency varied depending on the conditions, e.g., the shape of the function, initial sampling point, and hyper-parameters, but overall, the game tree search performed better than EI and UCB.}

\bibliography{swdbib}

\end{document}